# Field-Effect Transistors Based on Few-Layered α-MoTe$_2$

*Nihar R. Pradhan[1], Daniel Rhodes[1], Simin Feng[2], Yan Xin[1], Shahriar Memaran[1], Byoung-Hee Moon[1], Humberto Terrones[3], Mauricio Terrones[2], and Luis Balicas[1,*]*

[1]National High Magnetic Field Lab, Florida State University, 1800 E. Paul Dirac Dr. Tallahassee, FL 32310.

[2]Department of Physics, Department of Materials Science and Engineering and Materials Research Institute, The Pennsylvania State University, University Park, Pennsylvania 16802, USA

[3]Department of Physics, Applied Physics, and Astronomy Rensselaer Polytechnic Institute 110 Eighth Street, Troy, New York 12180-3590 USA.

**ABSTRACT**. Here we report the properties of field-effect transistors based on few layers of chemical vapor transport grown α- MoTe$_2$ crystals mechanically exfoliated onto SiO$_2$. We performed field-effect and Hall mobility measurements, as well as Raman scattering and transmission electron microscopy. In contrast to both MoS$_2$ and MoSe$_2$, our MoTe$_2$ field-effect transistors (FETs) are observed to be hole-doped, displaying on/off ratios surpassing $10^6$ and typical sub-threshold swings of ~ 140 mV per decade.

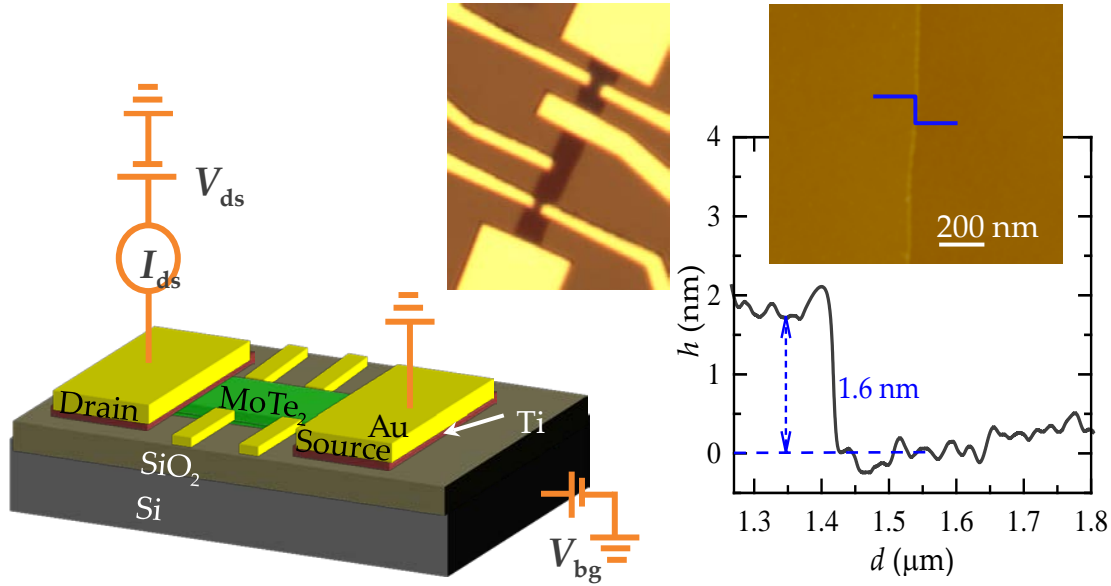

Both field-effect and Hall mobilities indicate maximum values approaching or surpassing 10 $cm^2$/Vs which are comparable to figures previously reported for single or bi-layered $MoS_2$ and/or for $MoSe_2$ exfoliated onto $SiO_2$ at room temperature and without the use of dielectric engineering. Raman scattering reveals sharp modes in agreement with previous reports, whose frequencies are found to display little or no dependence on the number of layers. Given that both $MoS_2$ is electron doped, the stacking of $MoTe_2$ onto $MoS_2$ could produce ambipolar field-effect transistors and a gap modulation. Although the overall electronic performance of $MoTe_2$ is comparable to those of $MoS_2$ and $MoSe_2$, the heavier element Te should lead to a stronger spin orbit-coupling and possibly to concomitantly longer decoherence times for exciton valley and spin indexes.

Several semiconducting transition metal dichalcogenides (TM$X_2$, where TM = Mo, W, Re, Nb *etc.*, and $X$ = S, Se, Te) crystallize in the layered trigonal prismatic structure characterized by a weak van der Waals inter-planar coupling, strong in-plane covalent bonding, lack of inversion symmetry, and a strong spin orbit coupling.[1] In single layered form, $MoS_2$ was found to be a direct gap semiconductor with energy gaps located at the *K* and *K'* points of the Brillouin zone. Both the highest valence bands and the lowest conduction bands are formed primarily from the Mo *d*-orbitals.[2] The large spin-orbit interaction splits the highest valence bands at the *K* (*K'*) points.[3-6] When exfoliated into thin atomic layers, this band splitting opens the possibility of tuning the position of the Fermi level (e.g. with a back gate voltage) towards the top spin-orbit split valence band having spin polarized carriers. This could lead to spin polarized carrier transport at the interface between a dielectric layer and the $TMX_2$, opening a new research frontier in spintronics.[7] Spin-orbit also leads to coupled valley and spin degrees of freedom due to the lack of inversion symmetry in monolayer $TMX_2$ and, this allows optical pumping of a single valley (and spin) with circularly polarized light.[8,9] In fact, each valley has a definite chirality arising from its strong pseudospin-orbit coupling[10] with the chiralities of each valley opposite to each other, due to the preservation of time-reversal symmetry. This leads to possible practical ways to differentiate the two valleys and to address them individually.[10-12] It has just been demonstrated that one can optically generate valley coherent excitons in monolayer semiconductor $WSe_2$.[13] Most importantly, the decoherence times of the exciton valley pseudospin were found to be comparable, or even slower, than the electron–hole recombination time suggesting a certain robustness of the "valley index" as a result of the coupled valley and spin degrees of freedom[11]. This suggests that similar to spins, the valley degrees of freedom could be effectively explored for novel optoelectronic devices. Since all these technologically promising effects rely on the size of the semiconducting gap and on the strong spin-orbit effect characteristic of the $TMX_2$ compounds, it justifies a systematic evaluation of their overall properties as a function of the atomic numbers of the composing atoms which tunes both the size of the semiconducting gap and the strength of the spin-orbit effect. Here, we chose to evaluate the properties of $\alpha$-$MoTe_2$, which is characterized by a smaller semiconducting gap in the bulk, i.e. (1.0±0.1) eV in contrast to (1.29±0.1) eV for $MoS_2$,[14] and contains the 5$p$ element Te which is considerably heavier (by a factor of ~3.5) with respect to the 3$p$ element S, leading to a larger splitting of the valence band, i.e. (238±10) meV in contrast to (161±10) meV for $MoS_2$.[14]

To date, field-effect transistors (FETs) based on transition-metal dichalcogenides (TMDs), $MoS_2$ in particular, were shown to display large transconductance as function of gate voltage, significant ratios between the "on" and "off" currents (or $I_{ON}/I_{OFF}$ >10$^8$) owing to their large band gap (~ 1.2 to 1.8 eV), and sharp conductance threshold, or "sub-threshold swings" as sharp as ~60 mV/decade, thus implying an enhanced gate control. Along with these electrical properties, the planarity of the monolayers based on transition-metal dichalcogenides makes their transistors viable candidates for low power applications.[13] Furthermore, transparency, flexibility, and relative inexpensiveness make them ideal for low cost electronics.

Besides the exploration of their remarkable optical properties, much of the current effort in the field of TMDs is devoted to the understanding of their fundamental electronic properties when exfoliated in single- or few-layers. For example, when mechanically exfoliated onto $SiO_2$, $MoS_2$ and $MoSe_2$ were found to be electron-doped compounds[15,16] in contrast to $WS_2$ and $WSe_2$ which were found to display ambipolar behavior.[17-19] This would make these latter compounds ideal for complementary digital logic applications. The combination of single-layer TMDs with dielectrics having a large dielectric constant κ such as $HfO_2$ or $Al_2O_3$ led to several reports[13,20,21,22] claiming a remarkable increase in carrier mobility thus making this architecture potentially useful for liquid crystals and organic light emitting displays.[13]

Here, we explore the electrical properties of a still less studied compound, i.e. $MoTe_2$ which is characterized by a smaller semiconducting gap (0.8 eV or 1.24 eV when in single-layered form) relative to its S (a 3$p$ element) or Se (a 4$p$ element) analogs. Similarly to $MoS_2$ or $MoSe_2$, single atomic-layers of $MoTe_2$ can be produced by using the mechanical exfoliation technique. Here, we focus on few-layered crystals mechanically exfoliated onto a 270 nm thick $SiO_2$ layer. Previous reports[23] found that the field-effect mobility of transition metal dichalcogenides TMDs tends to increase, reaching a maximum value when crystals are composed of approximately 10 atomic layers. A systematic study on the dependence of its properties as a function of the number of layers will be reported elsewhere, including crystals grown transport agents other than iodine. We show that the overall performance of a $MoTe_2$ based field-effect transistor is comparable to similar devices based on $MoS_2$ or $MoSe_2$. But an optical gap quite close in value to the one of Si and an enhanced spin-orbit interaction (since Te is a 5$p$ element) suggests that this compound might be particularly suitable for optoelectronic applications in a complimentary range of wave-lengths.

**RESULTS AND DISCUSSION**

Figure 1 shows a micrograph of one of our bi-layered $MoTe_2$ field-effect transistors built by mechanically exfoliated $MoTe_2$ onto thermally grown 270 nm $SiO_2$ on Si. $MoTe_2$ crystals were grown by chemical vapor transport using iodine as the transport agent (see Methods for details). This particular $MoTe_2$ crystal is composed of a bi-layered section spanning from current contact $I^+$ (bottom of the micrograph) up to current contact $I^-$, and a tri-layered section spanning from contact $I^-$ up to top contact $I^+$. As shown below, we observed slightly higher field-effect mobility for the tri-layered section when compared to the bi-layered one. Here, we focus on results collected on the bi-layered and tri-layered material as well as on a seven-layer crystal *at room temperature* with the goal of comparing field-effect and Hall-mobilities and also in order to evaluate the differences of MoTe$_2$ relative to other transition-metal dichalcogenide atomic layers such as MoS$_2$ and MoSe$_2$.

As depicted in Fig. 1 **a** the Hall voltage $V_H$, for this particularly sample, was measured through voltage contacts 1 and 3 under an applied external magnetic-field $H$, while the sample resistivity $\rho_{xx} = R_{xx}\, w/l$ ($w$ is the channel width and $l$ the separation of contacts) was measured through voltage leads 1 and 2 by using a conventional lock-in technique. The field-effect mobility $\mu_{FE} = (1/c_g)\, d\sigma/dV_{bg}$ ($c_g$ is the gate capacitance) is extracted by measuring the source-drain current $I_{ds}$ as a function of the back-gate voltage $V_{bg}$ under a constant source-drain excitation voltage $V_{ds}$ (therefore, $\sigma = I_{ds}/V_{ds}\, L/w$ is the two-terminal conductivity).

Figure 1 **b** shows an atomic force microscopy (AFM) image of the MoTe$_2$ single-crystal composing the FET shown in Fig. 1 **a**. The blue depicts the line along which the height profile shown in Fig. 1 **c** was acquired, which shows a step of ~ 1.6 nm; a value which is quite close to that reported for the MoTe$_2$ inter-layer spacing of $c = 1.396$ nm[24] and confirming thus that it is a bi-layered crystal. The inset in Fig. 1 **b** shows an atom resolved transmission electron-microscopy image of a vapor transport grown MoTe$_2$ single-crystal from the same synthesis batch used to fabricate the FET. For this image the electron beam flows perpendicularly to the atomic layers. In areas of the order of 100 nm$^2$, we could not observe any evidence for Te

vacancies or domain boundaries, thus indicating a high level of uniformity and crystallinity. This is also confirmed by diffraction pattern shown in the inset of Fig. 1 **c**, which reveals sharp Bragg peaks with no evidence for diffuse scattering.

Raman spectroscopy is a conventional and powerful tool able to identify the thickness of few-layered TMDs single-crystals[25-27]. In $MoS_2$ the $E_{2g}$ and $A_{1g}$ Raman modes were shown to be particularly sensitive with respect to the number of layers with their characteristic frequency shifting in opposite direction as the number of layers decreases.[25] Similar behavior were also identified in other TMDs such as $MoSe_2$ and $WSe_2$.[26,27] Here, we performed a not yet exhaustive, study on the evolution of the Raman modes of $MoTe_2$ as a function of the number of layers. Figure 2 **a** shows an AFM image of a $MoTe_2$ single crystal whose thickness varies from multiple atomic layers (outside of the scanned area), and becomes a bi-layer at its right bottom edge. This crystal was used by us to collect Raman spectra as a function of the number of layers. The blue line depicts the path along which the height profile shown in Fig. 2 **b** was acquired. The height profile starts at a height of ~ 1.7 nm, or approximately two atomic layers, and then increases by ~ 1 nm implying an additional atomic layer, or a triple-layered region as the AFM tip scans from left to right along the blue line in in Fig. 2 **a**. Figure 2 **c** depicts the Raman spectra for the bulk material (magenta line, multiplied by a factor of 5), and for the areas consisting of three- (blue) and two- layers (red), respectively. At first glance, and in contrast with previous reports on Raman spectra on other TMDs,[25-27] the main peaks would seem to not shift in frequency as the number of layers decrease. But by collecting Raman spectra at several points in each sample, and by averaging their position after a Lorentzian fit, one sees that the $A_{1g}$ peak red-shifts while the $E^I_g$ one (i.e. $E^1_{2g}$ mode in bulk crystals) blue shifts as the number of layers decrease, similarly to what is observed in other TMDs.[25-27] After the submission of this manuscript, we became aware of a complete Raman study on α-MoTe$_2$ as a function of the number of layers by Yamamoto *et al.*[28] Our results are qualitatively similar to those reported in this manuscript, however, and based on our ab-initio density functional theory (DFT) and density functional perturbation theory (DFPT) calculations described in Ref. 25 for MoS$_2$, MoSe$_2$ and WSe$_2$, we index the peak at ~290.9 cm$^{-1}$ as the $A^2_{1g}$ mode and not as the $B^1_{2g}$ as implied by Ref. 28. The un-indexed peak centered at 343.95 cm$^{-1}$ is observed to blue shift as the number of layers decreases. The un-assigned peak seems to be a second-order resonant peak and careful resonant Raman studies are needed to understand its nature.

Figure 3 **a** shows the extracted drain to source current $I_{ds}$ normalized by the width *w* of the channel as a function of the back-gate voltage $V_{bg}$ and for several values of the excitation voltage $V_{ds}$. As seen, for gate voltages surpassing ~ 20 to 30 V (depending on the value of $V_{ds}$), $I_{ds}$ remains at the level of the noise floor of our experimental set-up, hence MoTe$_2$ is a hole-doped compound. In effect, one has to reach positive gate values to "turn-off" the FET, thus indicating that this compound has an excess of holes in the conducting channel, in sharp contrast with most reports on MoS$_2$.[23] We verified that this behavior is intrinsic, i.e. it is observed under vacuum or helium atmosphere. Figure 3 **b** shows the same $I_{ds}$ as function of $V_{bg}$ but in a linear scale. The inset displays $I_{ds}$ as function of $V_{bg}$ again but in a limited range of current and gate voltages and for an excitation voltage $V_{ds}$ = 500 mV; from it one can extract a sub-threshold swing of ~140 mV. Figure 3 **c** plots the conductivity σ as function of $V_{bg}$ for each value of $V_{ds}$ in Figs. 3 **a** and 3 **b**. As seen, all curves for the different $V_{ds}$ values collapse on a single curve, indicating that this FET behaves linearly through the entire range of $V_{ds}$ excitations (see also the Supplementary Information section for leakage current as a function of back-gate voltage). The right panel of Fig. 3 **c** depicts the field-effect mobility $\mu_{FE}$ obtained, as previously mentioned, by taking the derivative of σ with respect to $V_{bg}$ and by normalizing it with the value of the geometrical gate capacitance of SiO$_2$, i.e. $c_g = \varepsilon_0 \varepsilon_r / d = 12.783 \times 10^{-9}$ F/cm$^2$ (where $\varepsilon_r = 3.9$ and $d$ = 270 nm is the SiO$_2$ thickness). $\mu_{FE}$ is seen to saturate at value of ~ 20 cm$^2$/Vs which is considerably higher than the initial reports for MoS$_2$ on SiO$_2$,[29] and even 2 orders of magnitude larger than the $\mu_{FE}$ values reported for bi-layered MoS$_2$ on SiO$_2$.[29] Although, it also surpasses the overall performance of multi-layered Sn(S,Se)$_2$,[31,32] these values still remain considerably lower, by two orders of magnitude, than those reported for multi-layered TMDs on SiO$_2$,[32,33] or for their single-layers in conjunction with high-κ dielectrics.[21,22] Figure 3 **d** shows $I_{ds}$ as a function of $V_{ds}$ for several values of the gate capacitance. As seen, the response is essentially linear, particularly at low values of $V_{ds}$ which is difficult to reconcile with a sizeable Schottky barrier $\phi_{SB}$ at the level of the current contacts. In the text below, we show an attempt to evaluate the size of $\phi_{SB}$.

Figure 4 shows results for two FETs fabricated on few layered α-MoTe$_2$ single-crystals of distinct thicknesses, i.e. respectively a tri-layer and a seven layers one as determined through atomic force microscopy. In effect Fig. 4 **a**, shows $I_{ds}$ as a function of $V_{bg}$ for a FET based on a tri-layered α-MoTe$_2$ crystal, and for several values of the excitation voltage $V_{ds}$. As seen, the sub-threshold swing still is sharp, but the threshold gate voltage required for current extraction increases up to ~ + 40 V suggesting a stronger level of hole dopants when compared to the bi-layered crystal. Notice that this FET also shows a current on to off ratio surpassing 10$^6$. Fig. 4 **b** shows $I_{ds}$ as a function of $V_{bg}$ in a linear scale, where the red line is a linear fit from which extract the field-effect mobility $\mu_{FE}$ ~ 20 cm$^2$/Vs, which is nearly twice the value previously obtained for the bi-layered FET. In order to further explore the dependence of $\mu_{FE}$ on $n$ or the number of layers, we show in Fig. 4 **c**, $I_{ds}$ as a function of $V_{bg}$ for a FET composed of seven layers. The same figure shows traces for increasing (red line) and decreasing (blue line) gate voltage sweeps, which do *not* reveal a sizeable hysteresis. Fig. 4 **d** displays $I_{ds}$ as a function of $V_{bg}$ in a linear scale from which we extract $\mu_{FE} \sim 27$ cm$^2$/Vs. Therefore, similarly to what was previously reported[34] for MoS$_2$, the mobility of α-MoTe$_2$ increases as the number of atomic layers increases up to $n \sim 10$. Nevertheless, this mobility value is lower than the ones extracted by us[34] and other groups[35] for multi-layered mechanically exfoliated MoS$_2$ but approaches the most recent values reported for chemical vapor deposited MoS$_2$.[36,37] Our energy dispersive spectroscopy (EDS) analysis (see Supplemental Information) reveals ~ 3% in Te excess and a similar amount of incorporated iodine. Although 3 % is close to the typical EDS quantitative error bars, the lower mobilities observed here, coupled to the high levels of hole-doping, does suggest that the material is not perfectly stoichiometric. A similar situation was recently reported for natural MoS$_2$, where its intrinsic electron-doping was attributed to the presence of S vacancies (~$10^{13}$ cm$^{-2}$) as observed through a detailed transmission electron microscopy study.[38] At the moment, we are still the effect of distinct chemical transport agents with the hope of synthesizing a more stoichiometric material, since Te and iodine have contiguous atomic numbers and very similar ionic radius, which could conduce to site disorder. Therefore, our results combined with those of Ref. 38, suggest that in TMDs a small chalcogenide deficiency or excess, leads to electron- or hole-doping, respectively.

We have characterized the temperature (*T*) dependence of the two-terminal conductivity $\sigma$ for the 7-layer α-MoTe$_2$ FET with the goal of understanding both the nature of the conduction mechanism and the quality of the contacts, or the size of the Schottky barrier at the contacts. Fig. 5 **a** shows $\sigma$ in a logarithmic scale and as a function of $T^{-1/3}$. As indicated by the red lines (linear fits), for $V_{bg} \leq -10$ V the expected expression for two-dimensional variable hopping conductivity (VRH), i.e. $\sigma \propto e^{-(T_1/T)^{1/3}}$, provides an excellent description for our $\sigma(T)$, indicating a relevant role for disorder within the conducting channel. Notice that VRH like conductivity was already previously reported for MoS$_2$ on SiO$_2$.[38,39] For $V_{bg} > -10$ V $\sigma$ is better described by $\sigma \propto e^{-(T_0/T)}$ as shown in Fig. 5 **b**, implying thermally activated nearest-neighbor hopping with $T_0$ ranging from 300 to 500 K. In effect, the slope of the linear fits (red lines in Fig. 5 **b**), which yields the value of $T_0$, decreases with increasing $V_{bg}$. This is consistent with an increase in the density of states near the Fermi level[40] $\varepsilon_F$ (since $T_0 = (k_B \varepsilon_F a)^{-1}$, where $a$ is the average distance between defects) as one moves away from the mobility edge by sweeping the gate voltage.

Once one has acquired the two-terminal conductivity $\sigma$ as a function of the $T$ one can proceed with another relevant exercise: to evaluate the size of the Schottky barrier for carrier conduction across the current contacts. In effect, and although Fig. 3 indicates that the conductivity $\sigma$, as measured through a two-terminal configuration, is linear on excitation voltage $V_{ds}$ when $V_{bg} > V_{bg}^t$, it was discussed in several reports that the conduction through the drain and source contacts is presumably entirely dominated by Schottky barriers.[23,41] In effect, a sizeable Schottky barrier $\phi_{SB} = (1.2 \pm 0.5)$ eV is expected as the difference in energy between the work function of Ti, or 4.33 eV, and the electron affinity of MoTe$_2$, or $(3.1 \pm 0.5)$ eV.[42] The linear, or apparent ohmic, regime would result from thermionic emission or thermionic field emission processes. In thermionic emission theory, the drain-source current $I_{ds}$ is related to $\phi_{SB}$ through the expression:

$$I_{ds} = AA^*T^2\exp(e\phi_{SB}/k_BT)$$

Where $A$ is the area of the Schottky junction, $A^* = 4\pi e m^* k_B^2 h^{-3}$ is the effective Richardson constant, $e$ is the elementary charge, $k_B$ is the Boltzmann constant, $m^*$ is the effective mass and $h$ is the Planck constant.[43] In order to evaluate the Schottky barrier at the level of the contacts, in the top panel of Fig. 5 **c** we plot $I_{ds}$ normalized by the square of the temperature $T^2$ as a function of $e/k_BT$ and for several values of the gate voltage. Red lines are linear fits from which we extract the value of $\phi_{SB}(V_{bg})$. Fig. 5 **d** shows $\phi_{SB}$ in a logarithmic scale as a function of $V_{bg}$, notice the small value of $\phi_{SB}(V_{bg} = 0 \text{ V}) \approx 27.5$ meV which is considerably smaller than the above estimated value exceeding 1 eV. At the moment we do not have a clear explanation for such a dramatic discrepancy; perhaps, the rather smaller values for $\phi_{SB}$ result from defect states relatively close to the Fermi level. Nevertheless, one has to be cautious with the extraction of the Schottky barrier through this approach, since two-terminal measurements contain contributions from both the contacts and the conduction channel which, as discussed above, would seem to undergo disorder-induced carrier localization, thus masking the true behavior of the conduction mechanism across the contacts. Based solely on the plots in Fig. 5 we cannot unambiguously identify the predominant conduction mechanism. Therefore the values of $\phi_{SB}(V_{bg})$ extracted here should be taken with caution: the very small values would imply a good band alignment between Ti and MoTe$_2$ which is difficult to conceive.

Figure 6 **a** shows the bare Hall signal, $R_{xy} = \Delta V_H/I = (V_H (+H) - V_H (-H))/2I$, where $V_H$ is the voltage measured through voltage contacts 1 and 3 (see Fig. 1 **a**) as a function of the applied magnetic field $H$ for several values of the gate voltage. From the linear fits (red lines) we extract the Hall constant $R_H = R_{xy}/H = 1/ne$, where $n$ is the two-dimensional density of carriers. Figure 6 **b** displays the sheet resistivity $\rho_{xx}$ (measured through contacts 1 and 2 in Fig. 1 **a**) as a function of $V_{bg}$. When $V_{bg}$ varies from -10 to -60 V, $\rho_{xx}$ decreases from ~ 550 kΩ to approximately 150 kΩ. The resulting Hall mobility $\mu_H = R_H/\rho_{xx}$ is shown in Fig. 5 **c**. We note that $\mu_H$ increases from approximately 6 cm$^2$/Vs to values approaching 11 cm$^2$/Vs. In addition, Fig. 6 **d** shows the extracted two-dimensional carrier density $n = 1/eR_H$ as a function of $V_{bg}$; it varies from ~ $5 \times 10^{12}$ to ~ $1 \times 10^{13}$ cm$^{-2}$. Red line represents a linear fit from which we extract the effective gate capacitance $c_g = ne/V_{bg} = (21 \pm 6)$ nF/cm$^2$, thus indicating that the actual gate capacitance could be twice as large as the geometrical value resulting from the thickness of the SiO$_2$ layer (or $c_g = 12.783$ nF/cm$^2$). A larger experimental value for $c_g$ can be understood as produced by spurious charges in the conduction channel. It is noteworthy that an effectively larger experimental value for the gate capacitance would easily explain the difference between the maximum values for the field-effect and the Hall-mobilities; by using the gate capacitance measured from the Hall-effect in the transconductance expression used to extract the field-effect mobility one would obtain basically the same value for both mobilities. This suggests that the electrical contacts and their respective Schottky barriers (which are rather small as indicated in Fig. 5 **d**) do not play a determinant role in limiting the carrier mobility in our α-MoTe$_2$ based field-effect transistors.

**Conclusions**

MoTe$_2$ grown by chemical vapor transport using iodine as the transport agent, when mechanically exfoliated onto SiO$_2$ and electrically contacted by using a combination of Ti and Au, behaves as a hole-doped semiconductor in contrast to MoS$_2$ and MoSe$_2$. Electron dispersive spectroscopy analysis indicates the possibility of a small amount of excess Te and possibly iodine incorporation which perhaps are the sources of hole-doping in this compound. If hole-doping resulted from Fermi level pinning associated with the metals used for the electrical contacts, it would suggest that Mo(Se$_{1-x}$Te$_x$)$_2$ or Mo(S$_{1-x}$Te$_x$)$_2$ alloys might become ambipolar, or alternatively that stacking MoTe$_2$ atomic layers onto MoSe$_2$/MoS$_2$ ones, might also lead to ambipolar field-effect transistors, suitable for *pn* junctions.[45-47] Perhaps such architecture might lead to new band gap materials[44] with novel optoelectronic properties. For example, at a given layer-layer distance the holes in one atomic layer could bound to the electrons on the other one in order to create bound excitons.[45] For all three Mo compounds a systematic study on the role of specific contact metals in pinning the Fermi level either closer to the valence or to the conduction band has yet to be performed to clarify the above issues. It is also noteworthy that our MoTe$_2$ FETs display $I_{ON}/I_{OFF}$ ratios also surpassing $10^6$ and sub-threshold swings of ~ 140 mV, which is comparable or superior to MoS$_2$ or MoSe$_2$ on SiO$_2$ in the absence of a high-κ dielectric layer. Field-effect mobilities indicate for few layered MoTe$_2$ exfoliated onto SiO$_2$, carrier mobilities ranging from ~ 20 cm$^2$/Vs in bilayers to values approaching 30 cm$^2$/Vs in seven layers. The Hall-effect on the other hand, indicates the presence of spurious charges in the conduction channel which certainly act as scattering centers, thus limiting the carrier mobility. Raman spectroscopy as a function of the number of layers indicates that the $A_{1g}$, and $E^1_g$ modes blue shifts and red shifts, respectively as the number of layers decrease, similarly to what was observed for other transition metal dichalcogenides. Therefore, Raman scattering can also be used in this compound in order to identify the number of layers of any given crystal.

In MoS$_2$ optically excited holes forming trions when bound to the photo generated excitons, were found to remain spin polarized in a given valley (for a time much longer than the trion lifetime), even in the presence of strong exchange mediated spin relaxation processes.[11] This is claimed to result from a particular robustness of the hole valley-index as the result of the coupled valley and spin degrees of freedom. In MoTe$_2$ it remains to be verified if the stronger spin-orbit/valley coupling, which was previously shown by photoemission experiments to lead to a larger valence band spin-splitting when compared to MoS$_2$,[14] would also lead to longer trion spin coherence-times, thus making this compound particularly appealing for new photonic and optoelectronic devices.[11]

**Methods**

MoTe$_2$ single-crystals were synthesized through chemical vapor transport using iodine as the transport agent. 99.999 % pure Mo powder, and 99.999 % pure Te pellets, were introduced into a quartz tube together with 99.999% pure I (5.8 mg of I per cm$^3$). The quartz tube was vacuumed and brought to 1150 $^0$C and held at this temperature for 1.5 weeks at a temperature gradient of ~100 $^0$C. Subsequently, it was cooled down to 1050 $^0$C at a rate of 10 $^0$C /h, followed by another cool down to 800 $^0$C at a rate of 2 $^0$C/h. It was held at 800 $^0$C for 2 days, and subsequently quenched in air. Bi-layered flakes of MoTe$_2$ were exfoliated from these single-crystals by using the ``scotch-tape" micromechanical cleavage technique, and transferred onto *p*-doped Si wafers covered with a 270 nm thick layer of SiO$_2$. For making the electrical contacts 90 nm of Au was deposited onto a 4 nm layer of Ti *via* e-beam evaporation. Contacts were patterned using standard e-beam lithography techniques. After gold deposition, the devices were annealed at 250$^0$ C for ~ 2 h in forming gas. This was subsequently followed by high vacuum annealing for 24 h at 120$^0$ C. Atomic force microscopy (AFM) imaging was performed using the Asylum Research MFP-3D AFM. Electrical characterization was performed by using a combination of sourcemeter (Keithley 2612 A), Lock-In amplifier (Signal Recovery 7265) and resistance bridges (Lakeshore 370) coupled to a Physical Property Measurement System. The Raman spectra were measured in a backscattering geometry using a 488 nm laser excitation wave-length. Sub-Angström aberration corrected transmission electron microscopy was performed by using a JEM-ARM200cF microscope. EDS was performed through field-emission scanning electron microscopy (Zeiss 1540 XB).

**ASSOCIATED CONTENT**

**Supporting Information**. Leakage current $I_{Leakage}$ flowing through the MoTe$_2$/SiO$_2$ interface as a function of back gate voltage $V_{bg}$, indicating that the drain to source current extract from a typical MoTe$_2$ FET is intrinsic to the interface. Typical EDS spectra of a chemical vapor transport grown MoTe$_2$ single crystal. This material is available free of charge *via* the Internet at http://pubs.acs.org.

**AUTHOR INFORMATION**

**Corresponding Author:** *balicas@magnet.fsu.edu

**Author Contributions**

DR synthesized MoTe$_2$ single crystals and characterized their stoichiometry through EDS. NRP fabricated the FETs and performed their characterization in conjunction with DR, and LB. NRP performed atomic force microscopy of the exfoliated flakes. SF performed Raman spectroscopy and photoluminescense measurements under the guidance of HT and MT. Transmission electron microscopy was performed by YX and DR. LB wrote the manuscript through contributions of all authors. All authors have given approval to the final version of the manuscript.

Funding Sources

**This work was supported by the U.S. Army Research Office MURI Grant No. W911NF-11-1-0362. The NHMFL is supported by NSF through NSF-DMR-0084173 and the State of Florida.**

*Conflict of Interest:* The authors declare no competing financial interest.

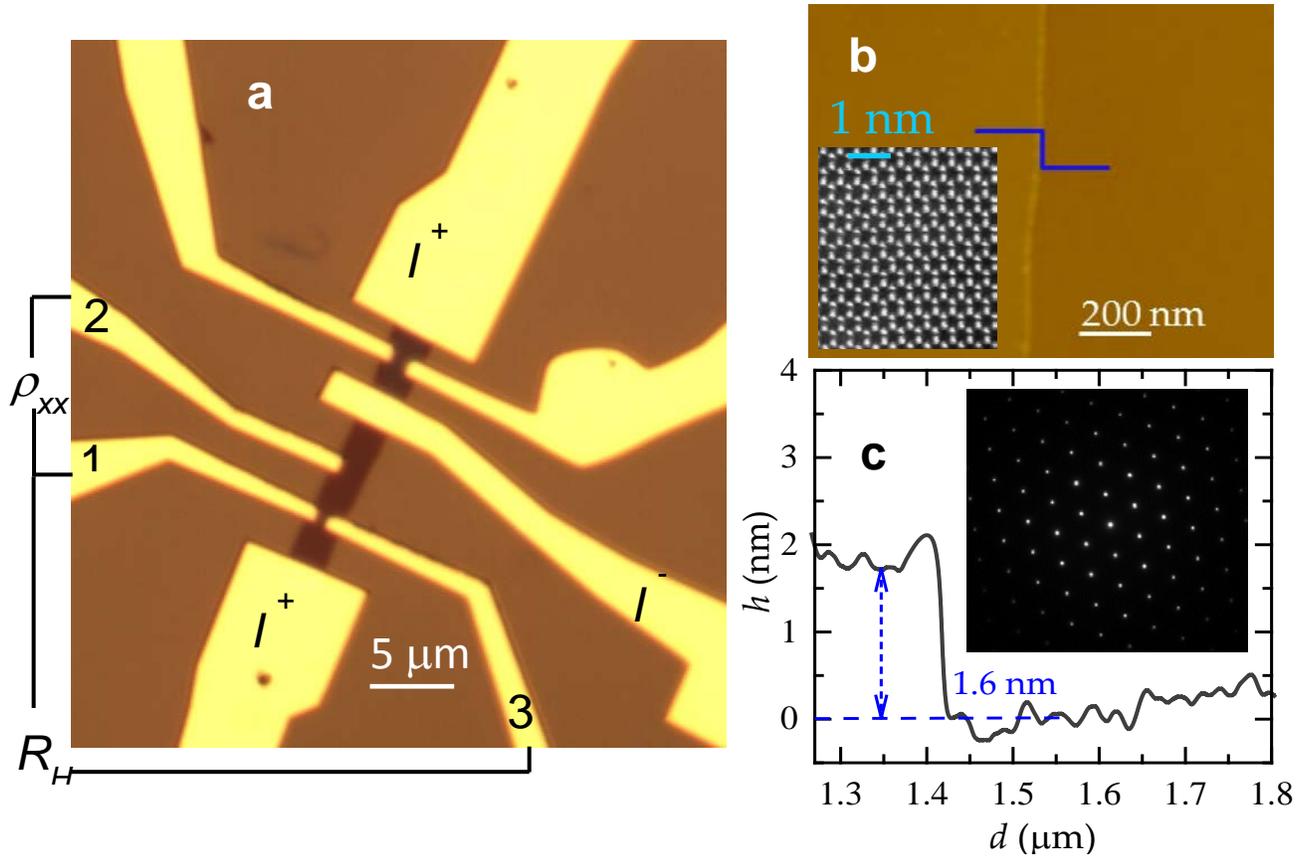

**Figure 1. a** Micrograph of one of our bi-layered MoTe$_2$ based field-effect transistors mechanically exfoliated onto SiO$_2$. The overall length $L$ of the channel between current contacts (labeled as $I^+$ and $I^-$, respectively) is $L = 10.847$ μm, while its width is $w = 4.172$ μm. The resistivity $\rho_{xx}$ is measured between voltage contacts 1 and 2 (separation $l = 1.55$ μm), while the Hall response $R_H$ was measured through the voltage contacts 1 and 3, with magnetic field applied perpendicularly to the plane of the figure. **b** The height of the flake $h$ is measured through an atomic force microscopy (AFM) scan long the edge of the sample (i.e. along the blue line). Inset: Transmission electron microscopy image showing the atomic positions within a plane perpendicular to the inter-layer direction. **c** AFM height profile along the blue line in **b**, indicating a flake height of ~ 16 Å which is quite close to its reported[22] interlayer distance c = (13.964 ± 0.004) Å thus indicating that this is a bi-layered flake. Inset: Electron diffraction pattern showing well defined Bragg peaks thus confirming the high level of crystallinity for the MoTe$_2$ single crystal.

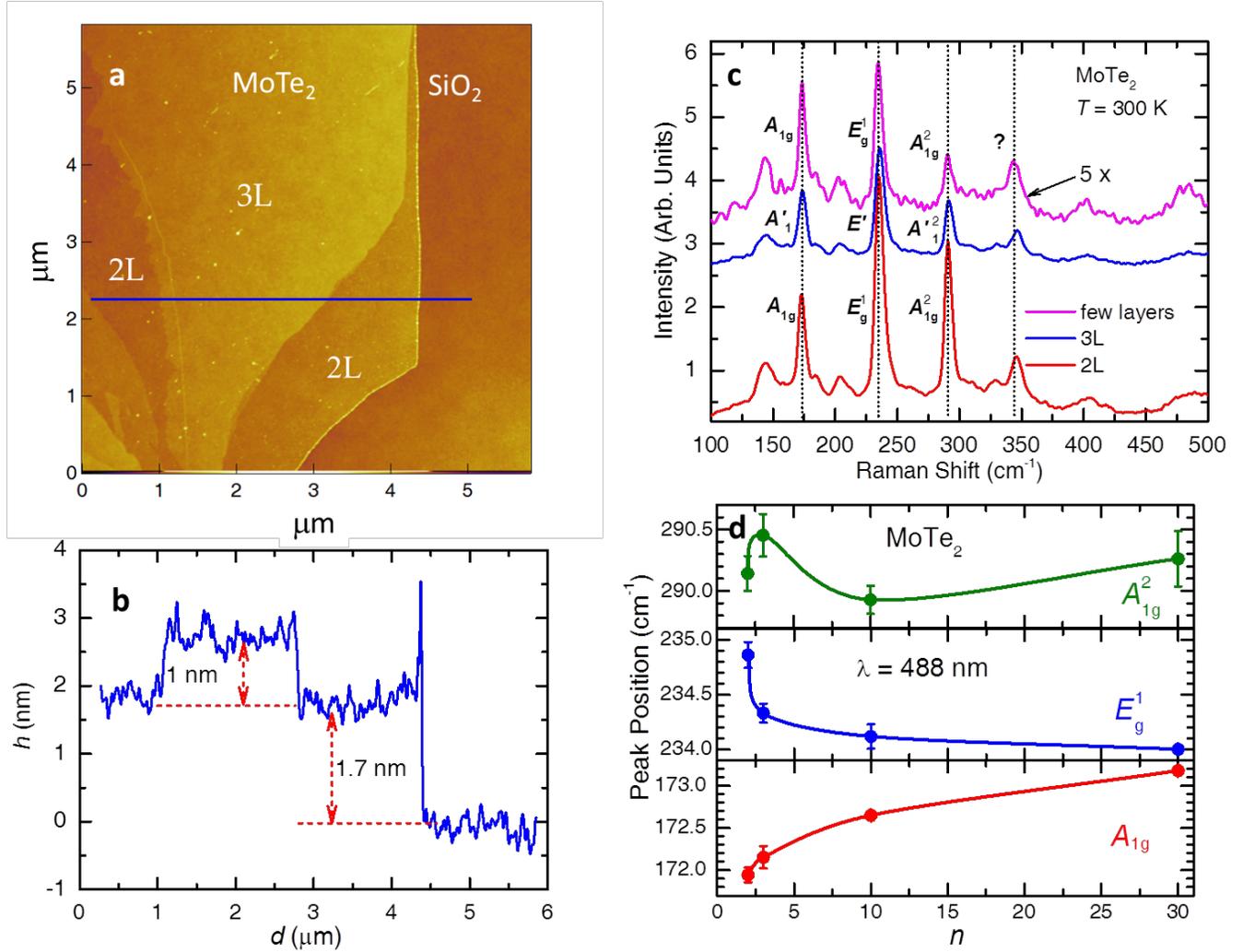

**Figure 2. a** Optical image of a MoTe$_2$ single crystalline flake, after mechanical exfoliation, used for Raman spectroscopy in conjunction with atomic force microscopy. As seen, the thickness of the flake varies, and as illustrated in **b** and **c**, certain areas of the flake are either bi- or tri-layered. The blue line in **b** indicates the line along which we acquired the height profile displayed in **c**, showing areas composed of three or two atomic layers. **d** Preliminary Raman spectrum as a function of the number $n$ of layers. The $A_{1g}$ and the $E^1_g$ modes were indexed based on previous Raman studies.[26] This preliminary study indicates that the position of the $A_{1g}$ mode (in cm$^{-1}$) red-shifts while the $E^1_g$ is seen to "blue shift" as the number of layers decrease. The extra sharp peaks observed in bilayers at 290.94 and ~ 346 cm$^{-1}$ respectively, are relatively close to the values of 285.9 and ~353 cm$^{-1}$ previously reported for bilayer MoSe$_2$ and indexed as the $E^1_{2g}$ and a $B^1_{2g}$ mode, respectively. Based on our calculations[26] we index the 290.94 cm$^{-1}$ peak as the $A^2_{1g}$ mode and leave the 353 cm$^{-1}$ one as a peak to be indexed by future work. Notice how this last one shifts considerably as the number of layers decrease. We confirmed that similar dependence on $n$ is also observed when using lasers with excitation wavelengths of 514 and 647 nm, respectively.

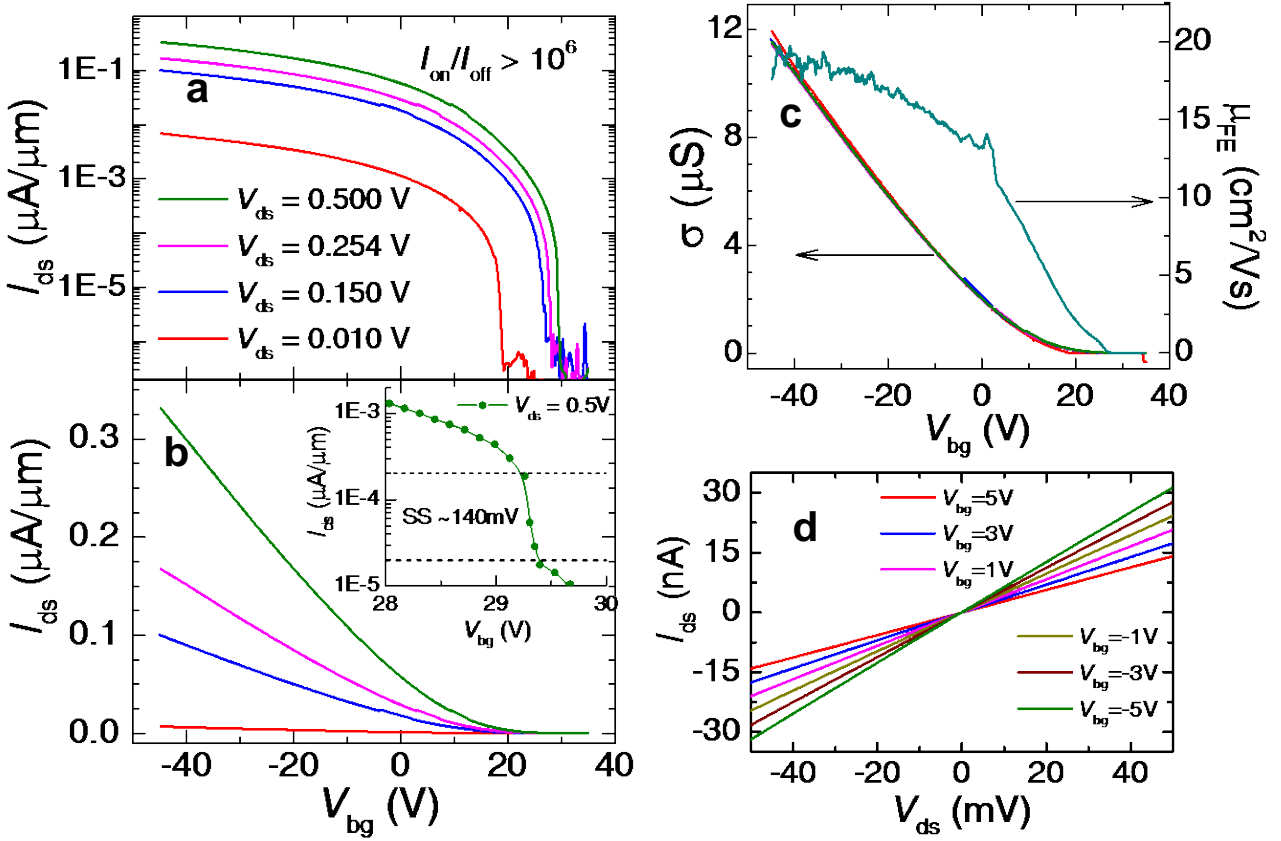

**Figure 3. a** Drain to source current $I_{ds}$ (in a logarithmic scale) as a function of the gate voltage $V_{bg}$ for a field-effect transistor based on bi-layered MoTe$_2$ and for several values of the drain-source excitation voltage $V_{ds}$. A sizeable current is observed for $V_{bg} < 0$ V, thus indicating that our synthetic MoTe$_2$ single-crystals are hole-doped. Notice also that the ratio of "on" to "off" field-effect current is $> 10^6$. **b** Same as in **a** but in a linear scale. Inset: $I_{ds}$ as a function of $V_{bg}$ in a limited scale, indicating a sub-threshold swing SS ~ 140 mV. **c** Right axis: conductivity $\sigma = SL/w$, where the conductance $S = I_{ds}/V_{ds}$ as a function of $V_{bg}$. Left axis: field-effect mobility $\mu_{FE} = (1/c)\,|d\sigma/dV_{bg}|$, where $c = \varepsilon_0\varepsilon_r/d$ is the gate capacitance ($d$ is thickness of the SiO$_2$ layer) as a function of $V_{bg}$. **d** $I_{ds}$ as a function of the excitation voltage $V_{ds}$ and for several values of $V_{bg}$. As seen, at low excitation voltages the $I$ - $V$ characteristics is linear.

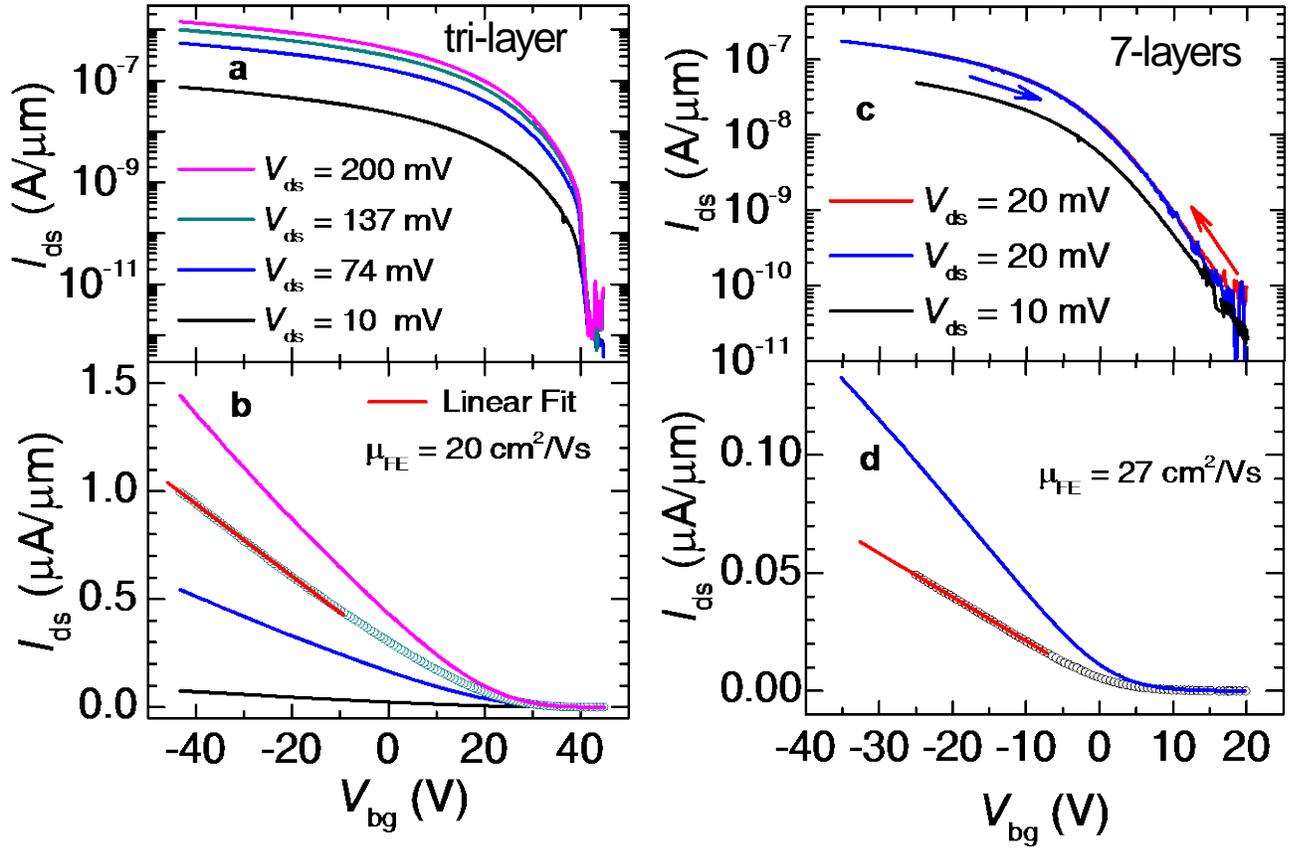

**Figure 4. a** Drain to source current $I_{ds}$ (in a logarithmic scale) as a function of the gate voltage $V_{bg}$ for a field-effect transistor based on a *tri-layered* MoTe$_2$ crystal and for several values of the drain-source excitation voltage $V_{ds}$. For this particular sample, a sizeable current is observed for $V_{bg} < 40$ V. Notice also that the on to off ratio still is $> 10^6$ while resulting field effect mobility $\mu_{FE} \sim 20$ cm$^2$/Vs is higher than the one observed for the bi-layered crystal. **b** Same as in **a** but in a linear scale. **c** $I_{ds}$ as a function of $V_{bg}$ for a field-effect transistor based on a *seven-layers* MoTe$_2$ single crystal and for two values of the drain-source excitation voltage $V_{ds}$. **d** Same as in **c** but in a linear scale. Red line is a linear fit from which we extract $\mu_{FE} \sim 27$ cm$^2$/Vs.

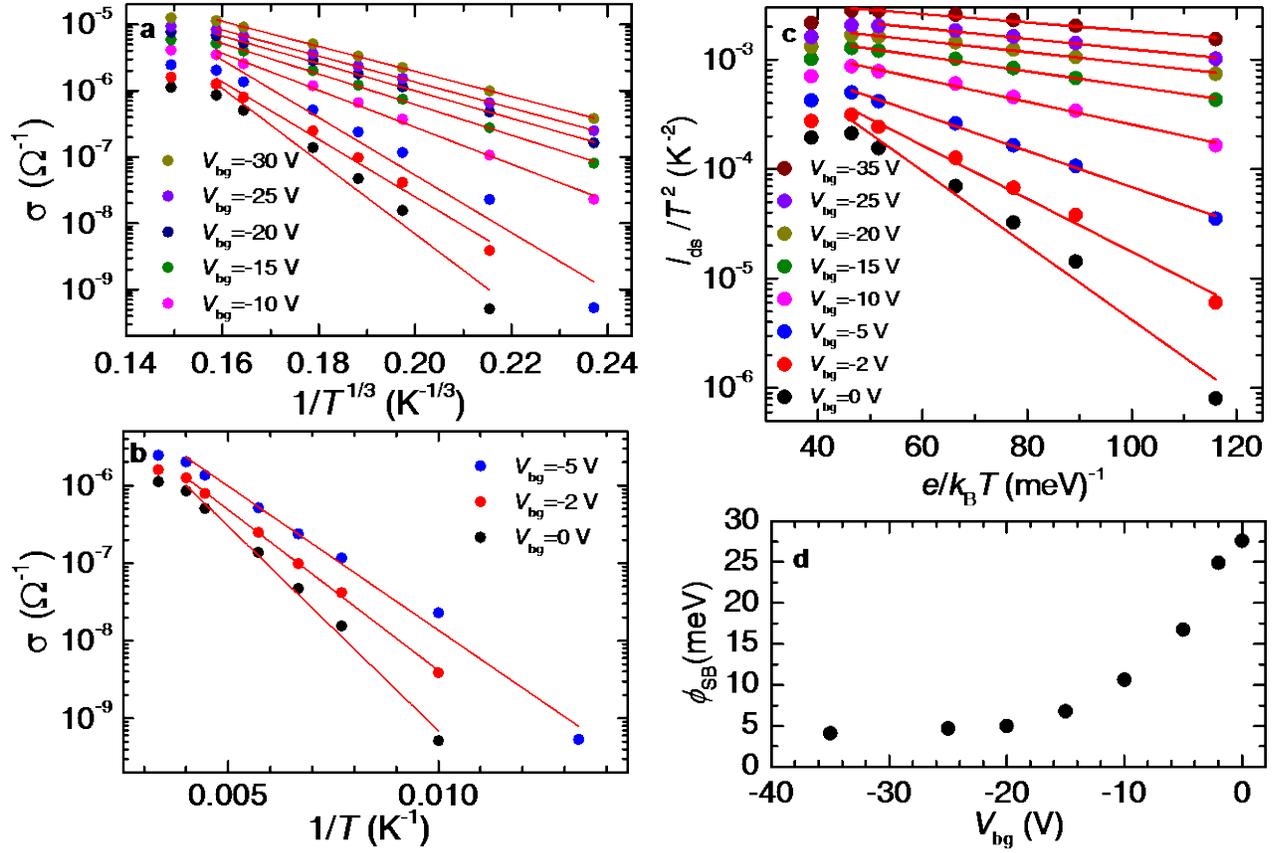

**Figure 5. a** Two-terminal conductivity $\sigma = I_{ds}/V_{ds}\, l/w$, where $l$ and $w$ are the length and the width of the conducting channel, respectively, in a logarithmic scale and as function of $T^{-1/3}$ for our *seven layers* sample and for several values of the back-gate voltage. Red lines are linear fits indicating that two-dimensional variable range hopping conductivity describes the behavior of $\sigma$ for $V_{bg} \leq -10$ V. **b** $\sigma$ in logarithmic scale as a function of $T^{-1}$, indicating that at low gate voltages, i.e. $-5\text{ V} \leq V_{bg} \leq 0\text{ V}$, $\sigma$ is better described by a behavior activated in temperature. **c** $I_{ds}$ normalized by the square of the temperature as a function $e/k_B T$. Red lines are linear fits, which according to the thermionic emission formalism would yield the size of the Schottky barrier $\phi_{SB}$ for carrier conduction across the current contacts. **d** $\phi_{SB}$ as a function of $V_{bg}$. Notice the quite small size of $\phi_{SB}$ ($V_{bg} = 0$ V).

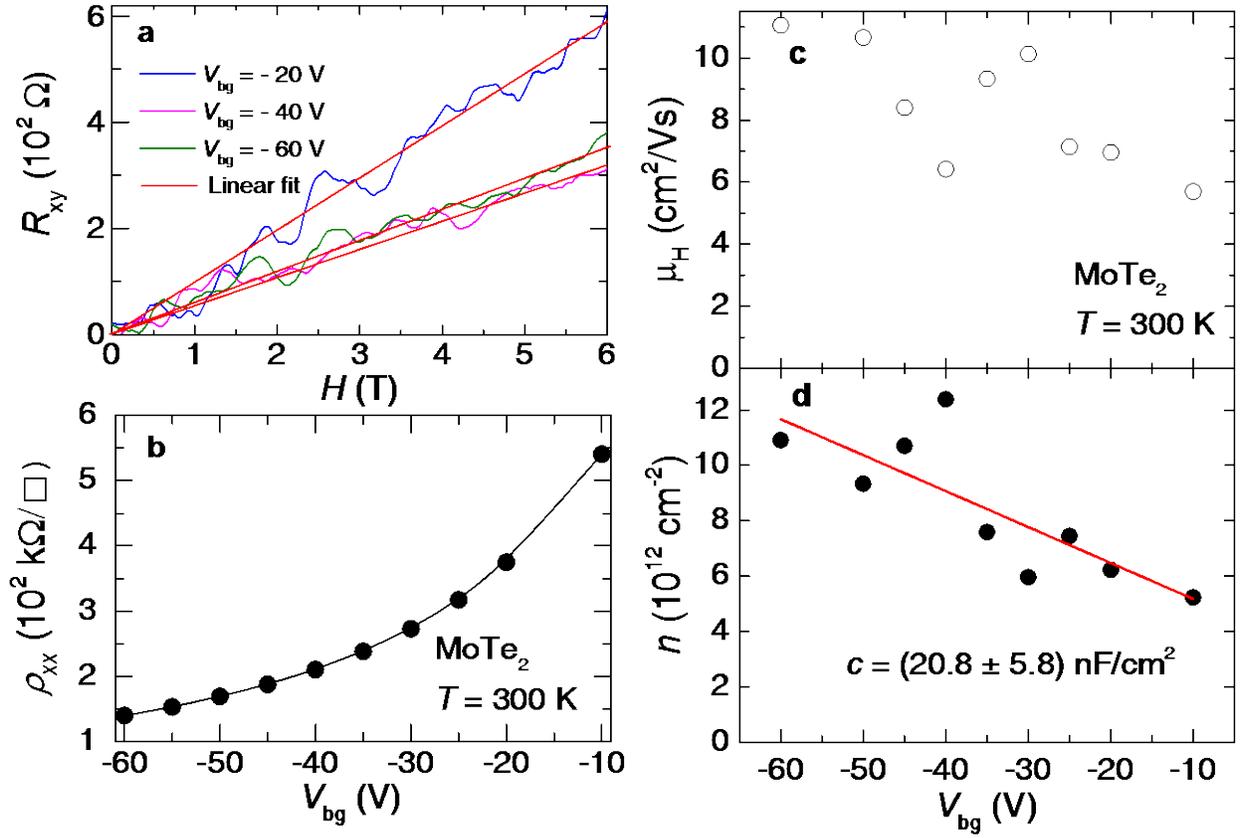

**Figure 6. a** Hall resistance $R_{xy}$ as a function of the magnetic field $H$ for a bi-layered $MoTe_2$ FET at room temperature. Red lines are linear fits from which we extract the Hall constant $R_H = 1/ne$ ($n$ is the density of carriers and $e$ is the electron charges). **b** Sheet resistivity $\rho_{xx}$ as a function of the back gate voltage $V_{bg}$. **c** Hall mobility $\mu_H = R_H/\rho_{xx}$ as a function of $V_{bg}$. **d** Carrier density $n = 1/eR_H$ as a function of $V_{bg}$. Red line is a linear fit from which one extracts the gate capacitance $c = ne/V_{bg}$. The extracted gate capacitance $c_g = (21 \pm 6)$ nF/cm$^2$ is higher than the expected value for a 270 nm thick $SiO_2$ layer, i.e. $c_g = 12.783$ nF/cm$^2$, indicating the presence of spurious charges in the conducting channel.

Supplemental Information For Manuscript Titled: **"Field-Effect Transistors Based on Few-Layered α-MoTe$_2$"** by *Nihar R. Pradhan[1], Daniel Rhodes[1], Simin Feng[2], Yan Xin[1], Shariar Memaran[1], Byoung-Hee Moon[1], Humberto Terrones[3], Mauricio Terrones[2], and Luis Balicas[1]*

[1]*National High Magnetic Field Laboratory, Florida State University, Tallahassee-FL 32310, USA*
[2]*Department of Physics, Department of Materials Science and Engineering and Materials Research Institute, The Pennsylvania State University, University Park, PA 16802, USA*
[3]*Department of Physics, Applied Physics, and Astronomy Rensselaer Polytechnic Institute 110 Eighth Street, Troy, New York 12180-3590 USA.*

**Leakage current as a function of the gate voltage**

Fig. 1 below shows a typical trace of the current flowing from the back-gate to the drain contact as a function of the back-gate voltage $V_{bg}$ at a fixed excitation voltage $V_{ds}$. Notice that the leakage current never surpasses 200 pA, even when the drain to source current surpasses 1 µA.

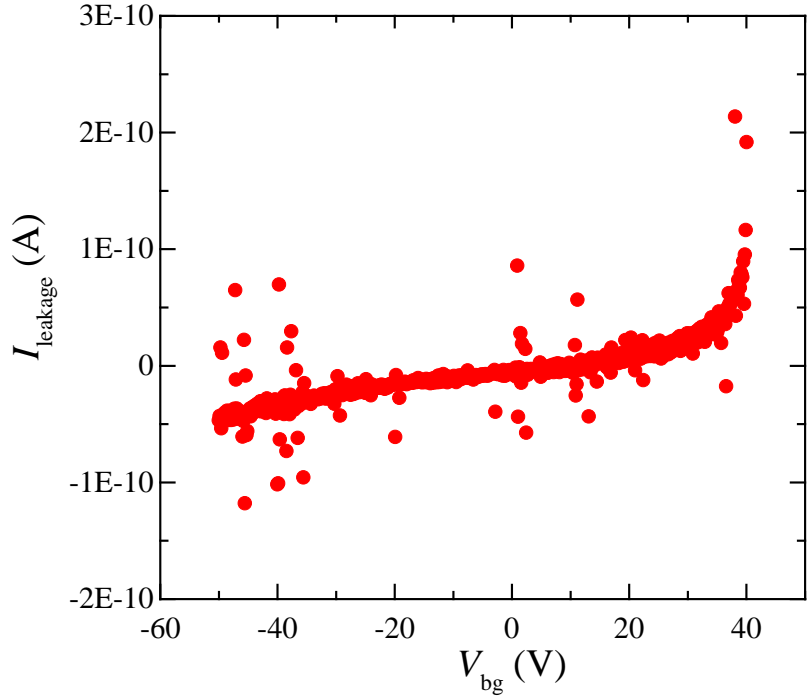

Fig. 1 Drain-source current $I_{ds}$ as a function of the drain-source excitation voltage $V_{ds}$ for several values of the back-gate voltage $V_{bg}$, at room temperature.

**Energy dispersion X-ray spectroscopy:**

Figure 2 below shows the energy dispersive x-ray spectroscopy collected on a single crystal in provenance of the same synthesis batch used for the fabrication of the field-effect transistors. As seen the only detected elements are Mo and Te and a fit of the spectra indicates a ratio very close to 1 to 2. Iodine was detected in other crystals and it is likely that an atomic number close to the one of Te leads to site/substitutional disorder.

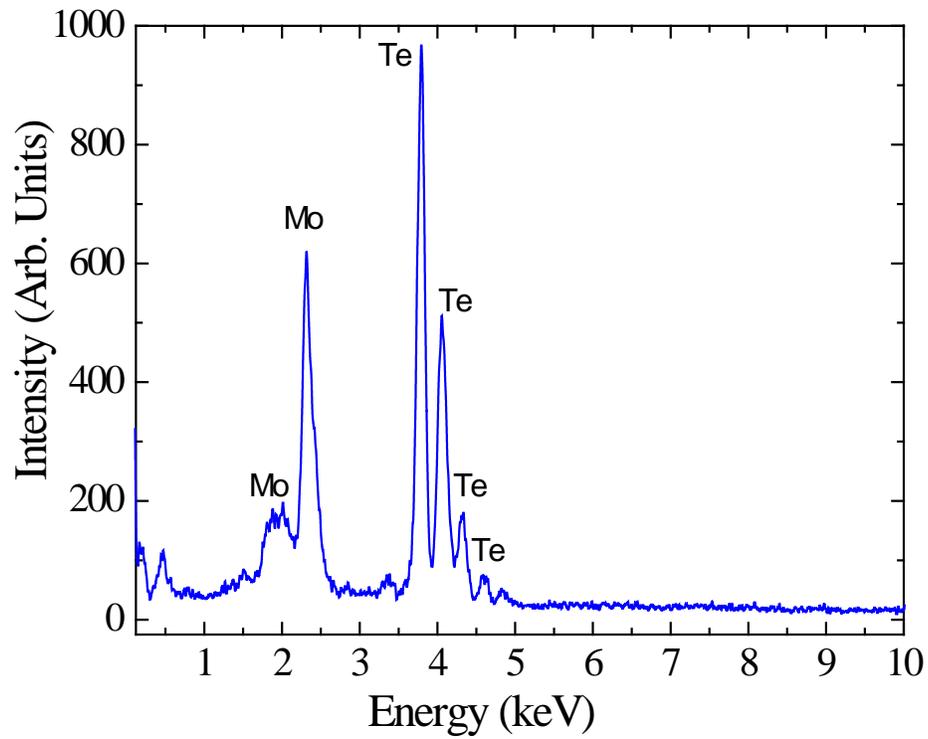

Fig. 2 Enery dispersive *x*-ray spectrum from a MoTe$_2$ single-crystal synthesized by chemical vapor transport method using iodine as the transport agent.